\documentclass[aoas,MSNbibl,nameyear,dvips]{arximspdf}
\usepackage{dcolumn}
\usepackage{graphicx}

\doi{10.1214/12-AOAS572} 
\volume{7}
\issue{1}
\pubyear{2013}
\firstpage{1}
\lastpage{24}

\makeatletter

\newcolumntype{d}[1]{D{.}{.}{#1}}

\newtheorem{theorem}{Theorem}

\makeatother

\begin{document}
\begin{frontmatter}

\title{Variance function estimation in quantitative mass spectrometry with
application to iTRAQ labeling}
\runtitle{Variance function estimation in quantitative MS}

\begin{aug}
\author[A]{\fnms{Micha}~\snm{Mandel}\corref{}\ead[label=e1]{msmic@huji.ac.il}},
\author[B]{\fnms{Manor}~\snm{Askenazi}},
\author[C]{\fnms{Yi}~\snm{Zhang}}\\
\and
\author[D]{\fnms{Jarrod~A.}~\snm{Marto}}

\runauthor{Mandel, Askenazi, Zhang and Marto}

\affiliation{Hebrew University of Jerusalem,
Dana-Farber Cancer Institute, Harvard Medical School and Hebrew
University of
Jerusalem, Dana-Farber Cancer Institute, and Dana-Farber Cancer
Institute and Harvard Medical School}

\address[A]{M. Mandel\\
Department of Statistics\\
Hebrew University of Jerusalem\\
Mount Scopus, Jerusalem\\
Israel, 91905\\
\printead{e1}} 
\address[B]{M. Askenazi\\
Departments of Cancer Biology\\
\quad and Blais Proteomics Center\\
Dana-Farber Cancer Institute\\
and\\
Department of Biological Chemistry\\
\quad and Molecular Pharmacology\\
Harvard Medical School\\
Boston, Massachusetts 02215-5450\\
USA\\
and\\
Department of Biological Chemistry\\
Hebrew University of Jerusalem\\
Jerusalem\\
Israel}
\address[C]{Y. Zhang\\
Departments of Cancer Biology\\
\quad and Blais Proteomics Center\\
Dana-Farber Cancer Institute\\
Boston, Massachusetts 02215-5450\\
USA}
\address[D]{J. A. Marto\\
Departments of Cancer Biology\\
\quad and Blais Proteomics Center\\
Dana-Farber Cancer Institute\\
and\\
Department of Biological Chemistry\\
\quad and Molecular Pharmacology\\
Harvard Medical School\\
Boston, Massachusetts 02215-5450\\
USA}
\end{aug}

\received{\smonth{10} \syear{2011}}
\revised{\smonth{5} \syear{2012}}

%
\begin{abstract}
This paper describes and compares two methods for estimating the
variance function associated with iTRAQ (isobaric tag for relative and
absolute quantitation) isotopic labeling in quantitative mass
spectrometry based proteomics. Measurements generated by the mass
spectrometer are proportional to the concentration of peptides present
in the biological sample. However, the iTRAQ reporter signals are
subject to errors that depend on the peptide amounts. The variance
function of the errors is therefore an essential parameter for
evaluating the results, but estimating it is complicated, as the number
of nuisance parameters increases with sample size while the number of
replicates for each peptide remains small. Two experiments that were
conducted with the sole goal of estimating the variance function and
its stability over time are analyzed, and the resulting estimated
variance function is used to analyze an experiment targeting aberrant
signaling cascades in cells harboring distinct oncogenic mutations.
Methods for constructing conservative $p$-values and confidence intervals
are discussed.
\end{abstract}

%

\end{frontmatter}

\section{Introduction} \label{secintro}

Improvements in mass spectrometer resolution, accuracy and
sensitivity coupled with the development of increasingly
sophisticated algorithms for protein identification from spectra
have resulted in mass spectrometry (MS) becoming the tool of choice in
large scale proteomics research. Typically, the mass spectrometer
is used to measure short portions of the proteins called peptides.
These are subjected to a process called tandem mass spectrometry (MS/MS)
which ultimately yields a mass spectrum containing peaks which
correspond to the primary amino acid sequence and enable the
identification of peptides. When samples are labeled with iTRAQ
(isobaric tag for relative and absolute quantitation) stable isotope
reagents, the MS/MS spectra also contain peaks at predefined masses,
whose intensities provide a relative measure of the peptide abundance
in a set of samples. A single experiment can yield
tens of thousands of spectra identifying thousands of peptides
belonging to thousands of proteins. Analysis of samples from different
sources, for example, cells expressing different mutations in a known
oncogene versus the wild-type (e.g., nontransforming) counterpart can
provide insight as to the mechanisms by which specific genetic lesions
manifested in the same protein drive a malignant phenotype. A workflow
diagram and a brief explanation of the iTRAQ technique are given in
Part A of the supplemental article [\citet{Manetal}]; for a
detailed discussion on MS techniques see \citet{EckObeThe}.

%
\begin{figure}

\includegraphics{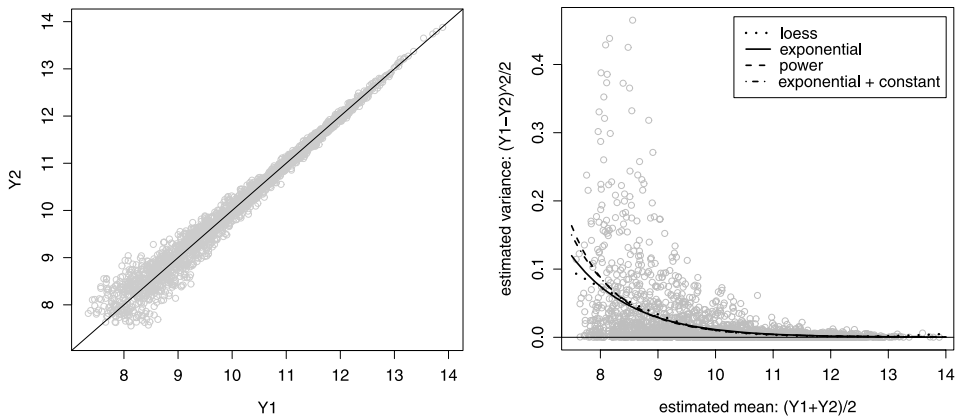}

\caption{Left: scatterplots of $Y_{i1}$ versus $Y_{i2}$
in the March experiment. Right: pointwise estimates of mean and
variance and different models for the variance.} \label{figdata}
\end{figure}

In an experiment conducted in March 2009 and described below,
\citet{Zhaetal10} applied the iTRAQ protocol using two different
labels for the same biological sample. The experiment yielded 2174
pairs of measurements corresponding to the amounts of 2174 peptides in
the sample. The left panel of Figure~\ref{figdata} depicts the
data\footnote{Pairs with exactly the same realized value were excluded
from the analysis; there were two such peptides.} (on a logarithmic
scale) and clearly shows that the variance of peak intensity
measurements is nonconstant and depends on the mean. This variance
should be estimated for better understanding of the results of the MS
analysis and to enable statistical inference about the peptide amounts.
The right panel of Figure~\ref{figdata} displays the mean of each pair
against its variance together with several estimates of the variance
function described in Section~\ref{secmix}.

The purpose of the current paper is twofold: first, to estimate the
variance function related to observations obtained by the common
technique of iTRAQ stable isotope labeling [\citet{Rosetal04},
\citet{AggChoLee06}], which provides measurements of the
relative amounts of peptides from two different biological samples in a
single experimental run; and second, to construct confidence intervals
for the abundance of a given peptide in a biological sample and to the
ratio of two abundances under different conditions (e.g., cancer and
wild type cells) using the estimated variance function, and to
calculate $p$-values for the hypothesis of equality in peptide amounts in
two independent samples.

Models that define the variance as a function of the mean have been
studied intensively in the framework of heteroscedastic regression,
where different estimation techniques have been suggested [e.g.,
\citet{DavCar87}]. However, in MS, the variance function depends
on the unknown peptide amount and the estimation problem is much more
involved. A similar problem arises in certain immunoassay studies where
few or no standard concentrations are available [\citet{Raa81},
\citet{SadSmi86}, \citet{OMaSmiSad08}], and in the evolving
area of microarray mRNA expression analysis [\citet{CarWan08},
\citet{WanMaCar09}, \citet{FanFenNiu10}]. Unlike immunoassay
and microarray, labeled MS data can contain as few as two measurements
of each peptide relative quantity and, therefore, analysis requires a
special experiment for estimating the variance function.

In a previous study, \citet{Zhaetal10} suggest a novel controlled
experiment for the estimation of the variance function in the iTRAQ
protocol. Using the standard workflow, they held the sample constant by
generating pairs of iTRAQ
intensities from identical biological samples. Under this experiment,
it is reasonable to assume that the iTRAQ labels are interchangeable in
their error characteristics since labeled samples are mixed before
processing and, therefore, the difference within pairs of such
measurements are entirely attributable to the measurement error of the
instrument itself. Two controlled experiments were conducted in January
and March of 2009 with the sole goal of estimating the variance
function [\citet{Zhaetal10}]. A separate experiment was conducted to
explore differences observed between wild-type and cancer cells
expressing distinct mutations of the same oncogenic kinase (FLT3); this
experiment relied on the variance function estimated in the controlled
study. The motivation for our current study arises from these past
experiments, and below we describe in detail the mathematical problem,
suggest statistical methods to tackle it and apply them to the three
data sets mentioned above.

In Section~\ref{secestim} we present a naive method for estimating the
variance function employed by Zhang et~al. and explore its validity. We
show that the method works well when the error terms are typically
small, as is the case in the instrument\vadjust{\goodbreak} explored by
\citet{Zhaetal10}, but may yield biased estimators for the
variance function in other situations. We then suggest an alternative
mixture model approach for estimating the variance function and prove
its consistency. In Section~\ref{secCI} we use the estimated variance
function for interval estimation of the ratio of peptides across two
different biological conditions, and we apply the method to iTRAQ-based
MS analysis that is intended to decipher the oncogenic potential of two
different clinically relevant mutations in the FLT3 receptor tyrosine
kinase. Section~\ref{secpval} discusses testing of the hypothesis that
the amounts of peptides in two biological samples are equivalent. The
properties of the estimation approaches are investigated in
Section~\ref{secsim} by simulation. Section~\ref{secdisc} completes the
paper with some remarks.

\section{Estimation of the variance function} \label{secestim}

\subsection{The model} \label{secmodel}

Consider a controlled iTRAQ experiment that quantifies $N$ peptides.
Let $Y_{i1},Y_{i2}$ denote two measures of intensity of peptide
$i$ (on a logarithmic scale) having an unknown mean
$\mu_i$. Assume that $(Y_{i1},Y_{i2})$ ($i=1,\ldots,N$) are
independent following the model:
%
%
\begin{equation}
\label{varmodel} Y_{ij} \sim N\bigl(\mu_i,h(\theta,
\mu_i)\bigr),\qquad j=1,2,\qquad\mbox{independent},
\end{equation}
where $\theta$ is a vector of unknown variance parameters, and
$h(\theta,\mu)$ is a known positive function, such as the power
function $\theta_1\mu^{\theta_2}$ or the exponential function
$e^{\theta_1+\theta_2\mu}$. In problems where $\mu_1,\ldots,\mu_N$
are known or are modeled by a small number of auxiliary
variables, standard techniques for estimating heteroscedastic
regression models apply [e.g., \citet{DavCar87}].
However, this is not the case in MS data where
$\mu_1,\ldots,\mu_N$ are unknown nuisance parameters.

\citet{KlaHunJan06} and \citet{Hunetal09}
developed an EM algorithm that maximizes over $\mu_1,\ldots,\mu_N$ and
$\theta$ the likelihood corresponding to $N$ independent observations,
$(y_{i1},y_{i2})$, from model~(\ref{varmodel}):
%
%
\begin{equation}
\label{eqlik} \prod_{i=1}^N \bigl\{2\pi h(
\theta,\mu_i)\bigr\}^{-1}\exp\biggl\{-\frac
{(y_{i1}-\mu_i)^2+(y_{i2}-\mu_i)^2}{2h(\theta,\mu_i)}
\biggr\}.
\end{equation}
However, since the number of nuisance parameters increases with sample
size, the maximum likelihood approach may provide biased estimators.
The bias is readily seen in the classical example by
\citet{NeySco48} of the one-parameter homoscedastic model
$h(\theta,\mu)=\theta$. The maximum likelihood estimator for $\theta$
under this model is $N^{-1}\sum_i(Y_{i1}-Y_{i2})^2/4$ having
expectation $\theta/2$, hence it converges to half the true variance.
Thus, although the model is parametric, alternatives to the maximum
likelihood technique should be employed.

\subsection{The MACL approach of Sadler and Smith}
\label{secmacl}

Motivated by immunoassay data, \citet{Raa81} suggests to modify the
likelihood (\ref{eqlik}) by multiplying the contribution\vadjust{\goodbreak} of each of
the pairs
$(Y_{i1},Y_{i2})$ by $h^{1/2}(\theta,\mu_i)$, and then to maximize
the modified likelihood with respect to
$\mu_1,\ldots,\mu_N$ and $\theta$. Raab shows by simulation that the
standard maximum likelihood estimator is biased, but the modified
estimator performs reasonably well. Raab's method is computer
intensive, as it requires the estimation of all nuisance parameters.

\citet{SadSmi86} estimate $\theta$ by maximizing Raab's modified
likelihood at the point
$\mu_i=\bar{Y}_i:=(Y_{i1}+Y_{i2})/2$ $(i=1,\ldots,N)$:
%
%
\begin{equation}
\label{eqgenprb} \hat{\theta} = \mathop{\arg\max}_\theta\prod
_{i=1}^N \frac{1}{{4\pi^2
\sqrt{
h(\theta,\bar{Y}_i)}}} \exp\bigl
\{-S^2_i/2h(\theta,\bar{Y}_i)\bigr\},
\end{equation}
where $S_i^2:=(Y_{i1}-\bar{Y}_i)^2+(Y_{i2}-\bar
{Y}_i)^2=(Y_{i1}-Y_{i2})^2/2$. This approach, called maximum
approximate conditional likelihood (MACL) by Sadler and Smith, reduces
the estimation task to solving a small set of nonlinear equations.
%
%
The resulting
estimating equations under the normal model are therefore
\[
\sum_{i=1}^N \frac{ {\partial}
h(\theta,\bar{Y}_i)/{\partial\theta_j }}{h^2(\theta,\bar{Y}_i)} \bigl
\{S_i^2 - h(\theta,\bar{Y}_i) \bigr\} = 0.
\]
For example, for the model $h(\theta,\mu)=\exp(\theta_1+\theta_2
\mu)$, the estimating equations reduce to
%
%
\begin{eqnarray} \label{eqee1}
1 - N^{-1}\sum_{i=1}^N
S_i^2\exp(-\theta_1-\theta_2
\bar{Y}_i) & = & 0,
\\
\label{eqee2}
N^{-1}\sum_{i=1}^N
\bar{Y}_i- N^{-1}\sum_{i=1}^N
\bar{Y}_i S_i^2\exp(-\theta_1-
\theta_2 \bar{Y}_i) & = & 0,
\end{eqnarray}
and can be easily solved by standard optimization algorithms using, for
example, the R function \textit{optim} [\citet{Dev11}]. As pointed
out by Sadler and Smith, the solution for (\ref{eqgenprb}) can be
obtained by an iterative reweighted least squares algorithm.

We applied the MACL approach to the January ($N=2144$) and March
($N=2174$) experiments described in Section~\ref{secintro} using the
functional form $h(\theta,\mu)=\exp(\theta_1+\theta_2\mu)$ and obtained
the estimates $\hat{\theta}=(4.89,-0.935)$ and $\hat{\theta
}=(4.89,-0.925)$, respectively.\footnote{Data and R codes can be
accessed as project syn310406 on the Sage Bionetworks Synapse system
(\url{http://synapse.sagebase.org}).} The similarity of the two estimated
variance functions is
remarkable, suggesting that the between-study variability is
small. This is a very important finding, as the variance function can
be estimated in a control experiment and be applied to data obtained in
independent experiments on the same instrument. We finally pooled the
data together and obtained an overall MACL estimate of $\hat{\theta
}=(4.86,-0.927)$.

Neither Raab nor Sadler and Smith provide sound
theoretical justification for their methods, but explore them in
several special relevant cases.\vadjust{\goodbreak}
In general, the expectation of the estimating equations differs from 0,
hence, the estimators of the variance function are, in general,
inconsistent. This is shown in Appendix~\ref{appA} and is illustrated
by simulation in Section~\ref{secsim}. However, Appendix~\ref{appA}
suggests that the bias is small when the variance (for all $\mu_i$) is
small, because in such circumstances $\bar{Y}_i$ is a good estimator
for $\mu_i$, even though it is based on only two observations.
Recently, \citet{WanMaCar09} studied variance functions for microarray
experiments and showed a similar inconsistency problem of estimators
obtained by the method of moments.

\subsection{A mixture model} \label{secmix}

A possible strategy to deal with the inconsistency of the
MACL approach is to impose additional reasonable
assumptions on the nuisance parameters. We consider the model
%
%
\begin{eqnarray}
\label{eqmixmod} %
Y_{ij}|
\mu_i &\sim& N\bigl(\mu_i,h(\theta,\mu_i)
\bigr),\qquad j=1,2,\qquad \mbox{independent},
\nonumber\\[-8pt]\\[-8pt]
\mu_i&\sim& G_0,\qquad i=1,\ldots,N,\qquad \mbox{independent},\nonumber
\end{eqnarray}
where (i) the support of $G_0$ is in the segment $[a,b]$, that is,
$P(a\le\mu_i \le b)=1$, (ii) the variance is bounded, that is,
$\alpha
<h(\theta,\mu)<\beta$ for all $\mu$ in the support of $G_0$ for
some $0<\alpha<\beta<\infty$, (iii) $h(\theta,\mu)$ is continuous on
$[a,b]$ and identifies $\theta$, that is, knowing $h(\theta,\mu)$ on
the support of $G_0$ implies knowledge of $\theta$.

The assumptions on $h$ are satisfied by most practical models. The
reason for bounding $G_0$ and the choice of the values $a$ and $b$ are
discussed in Section~\ref{secCI}. To see the importance of the
identifiability assumption, consider the model $h(\theta,\mu)=\exp
(\theta_1+\theta_2\mu)$ and a degenerate $G_0$ that assigns all the
mass to some $\mu_0$. In such a model, $\exp(\theta_1+\theta_2\mu_0)$
is identifiable, but the pair $(\theta_1,\theta_2)$ is not.

%
\begin{theorem}\label{theo1}
Under model (\ref{eqmixmod}) and the assumptions following it, the
maximum likelihood estimator of $(\theta,G_0)$ is consistent.
\end{theorem}

The proof, which is sketched in Appendix~\ref{appB}, is based on the seminal
paper by \citet{KieWol56} who prove the consistency of the maximum likelihood estimator in
mixture models such as (\ref{eqmixmod}). A recent application of
mixture models for variance
estimation in microarray analysis can be found in \citet{WanMaCar09},
though they suggest a different estimation strategy for $\theta$.

Several algorithms for deriving the maximum likelihood estimator of a
mixture model have been suggested in the literature [see, e.g.,
\citet{Boh99}]. Here we estimate $\theta$ and $G_0$ by employing
the EM algorithm, treating $(Y_{i1},Y_{i2},\mu_i)$ as the complete data
on the $i$th peptide. One strategy for estimating $G_0$ is to restrict
the search to distributions supported on a fine grid and to find the
maximum likelihood among these distributions. In the current problem,
the variance becomes small for large values of $\mu$ and using a simple\vadjust{\goodbreak}
grid may lead to data points which are too far (in terms of standard
deviations) from all support points. We found that defining the support
points of $G_0$ as a function of the variance performed better than
using a simple grid. Thus, we first obtained an initial estimate of the
variance function using, for example, the MACL approach, and then
restricted the distance between two support points to be at most $d$
standard deviations according to the estimated function. Specifically,
let $\tilde{\theta}$ be an initial estimate for $\theta$, and define
the maximal support point to be $\mu_J=b$, the second to maximal to be
$\mu_{J-1}=\mu_J-d\sqrt{h(\tilde{\theta},\mu_J)}$ and recursively
$\mu_{j-1}=\mu_j-d\sqrt{h(\tilde{\theta},\mu_j)}$. The selected points
of support depend on the initial estimate of the variance function and
can be updated as part of the algorithm, though in our experience, this
update made no significant improvement when using $d=1/4$. Once the
support points for $G_0$ are determined, the EM algorithm is applied to
estimate $\theta$ and $G_0$. The algorithm is quite standard and its
description is detailed in Appendix~\ref{appC}.


We fit the following three forms for $h(\theta,\mu)$ that have been
suggested in the literature\footnote{Data and R codes can be accessed
as project syn310406 on the Sage Bionetworks Synapse system
(\url{http://synapse.sagebase.org}).}: $\exp(\theta_1)\mu^{\theta_2}$,
$\exp
(\theta_1+\theta_2\mu)$ and $\exp(\theta_1+\theta_2\mu)+\exp
(\theta_3)$.
The right panel of Figure~\ref{figdata} presents estimates of these
three functions on a scatter diagram of $\bar{Y}_i=(Y_{i1}+Y_{i2})/2$
against $S^2_i=(Y_{i1}-Y_{i2})^2/2$. The X-axis is a naive estimate of
the mean of each pair, $\mu_i$, and the Y-axis is a naive estimate of
the variance. All three models give similar results in most of the
range with some deviation for very small values of $\mu$. A loess fit,
presented in the figure by the dotted line, is quite close to the
functional form $h(\theta,\mu)=\exp(\theta_1+\theta_2\mu)$ originally
suggested by \citet{Zhaetal10}. This latter model is used throughout
this paper.

Applying the EM approach using the functional form $h(\theta,\mu
)=\exp
(\theta_1+\theta_2\mu)$, we obtained the estimates $\hat{\theta
}=(4.91,-0.929)$ and $\hat{\theta}=(4.96,-0.944)$ based on the January
and March experiments, respectively. The estimates are very similar to
the estimates obtained by the MACL approach. We then pooled the data
from the two experiments together and obtained our final estimate,
$\hat
{\theta}=(4.84,-0.927)$. The corresponding estimate of $G_0$ is
displayed in the supplemental article [\citet{Manetal}].
Initial values for the EM algorithm were obtained by the MACL approach.
Starting the algorithm from different points resulted in essentially
the same estimate (details are provided in the supplementary materials).

\section{Confidence intervals} \label{secCI}

Having estimated the variance function, confidence intervals for
various parameters of interest can be constructed based on data from a
new experiment that compares different biological samples.\vspace*{1pt}
We construct frequentist confidence intervals that are based on the
estimate $\hat {\theta}$ of $\theta$ under model (\ref{eqmixmod}), but
do not use the estimate of the mixing distribution $G_0$. We take this
approach since the variance function is a stable characteristic of the
mass spectrometer, whereas $G_0$ depends on the biological sample and
may differ from sample to sample.

\subsection{\texorpdfstring{Confidence intervals for $\mu$}{Confidence intervals for mu}} \label{secCI1par}

It is of interest to attach a measure of uncertainty to the observed
intensity or to report a range rather than one value for each peptide.
This section discusses the construction of $1-\alpha$ confidence
intervals for the peptide amount, $\mu$, based on one
observation $Y$ from the model $Y\sim
N(\mu,h(\theta,\mu))$; the next section deals with the
construction of confidence sets for the relative abundance of a peptide
in two different samples.

Similar to the MACL approach for estimation, construction of confidence
intervals can be simplified by plugging an estimate of $\mu$ in
$h(\theta,\mu)$. Thus, a naive confidence interval is constructed by
$Y\pm z_{1-\alpha/2}\sqrt{h(\hat{\theta},Y)}$, where $z_\alpha$ is the
$\alpha$ quantile of the standard normal distribution. This method is
expected to perform reasonably well only in cases were the variance is
small and changes slowly with $\mu$.

An exact $1-\alpha$ confidence set for $\mu$ can be constructed using
the pivotal
quantity $g_Y(\mu)=(Y-\mu)^2/h(\theta,\mu)$. This interval is
defined as $C_\alpha= \{\mu\dvtx g_Y(\mu)\le\chi_{1,1-\alpha}\}$,
where $\chi_{df,\alpha}$ is the $\alpha$ quantile of the chi-squared
distribution with $df$ degrees of freedom.

For the model $h(\theta,\mu)=\exp(\theta_1+\theta_2\mu)$, this
set can be
easily found by a bisection search using the following
observations:
%
\begin{longlist}[(2)]
\item[(1)]$g_Y(Y)=0$ is a local minimum.
\item[(2)] For\vspace*{1pt} $\theta_2<0$,
$\mu^*=Y+2/\theta_2$ is a local maximum of $g_Y(\mu)$, with
$g_Y(\mu^*)=4\theta^{-2}_2 e^{-(2+\theta_1+\theta_2Y)}$.
\end{longlist}
We thus obtain the following properties of the confidence procedure:
\begin{itemize}
\item If $g_Y(\mu^*)\le\chi_{1-\alpha}$, then the confidence set
is a one-sided interval of the form $(-\infty,r_1)$.

\item If $g_Y(\mu^*)> \chi_{1-\alpha}$, then the confidence set is
a union of two intervals, $(-\infty,r_1)\cup(l_2,r_2)$, where
$r_1<\mu^*<l_2<Y<r_2$.
\end{itemize}
This nonstandard shape of the confidence set reflects the fact
that a realization $y$ is likely either when $\mu$ is close to $y$
or when $\mu$ is much smaller than $y$ and the variance of the
measurement is
very large. Often, the range of $\mu$ is a priori bounded so that
values smaller than $r_1$ do not belong to the parameter space,
and the confidence set is always an interval.

Since the parameter $\theta$ is unknown, a consistent estimator based
on the mixture approach is plugged in
to generate confidence intervals with an
approximate level $1-\alpha$.

\subsection{\texorpdfstring{Confidence intervals for $\mu_1-\mu_2$}{Confidence intervals for mu1-mu2}} \label{secCI2par}

Let $Y_1\sim N(\mu_1,h(\theta,\mu_1))$ and $Y_2\sim
N(\mu_2,h(\theta,\mu_2))$ be the log intensities of the
same peptide obtained by the iTRAQ protocol under two different
conditions, and assume that $Y_1$ and
$Y_2$ are independent conditionally on $\mu_1$ and $\mu_2$. We consider
the construction of a
confidence set for $\mu_1-\mu_2$, which is the parameter of
primary interest in many quantitative MS studies.

As in the one-parameter case, a naive confidence interval for
$\mu_1-\mu_2$ can be obtained by plugging $Y_1$ and $Y_2$ into the
variance term: $Y_1-Y_2 \pm
z_{1-\alpha/2}\sqrt{h(\hat{\theta},Y_1)+h(\hat{\theta},Y_2)}$.
However, such intervals may be anti-conservative, as the estimator
$Y_i$ for $\mu_i$ is inconsistent, and hence so is the estimator
$h(\hat{\theta},Y_i)$ for $h({\theta},\mu_i)$.

A direct and a relatively simple way of calculating conservative
intervals is by Bonferroni correction, that is, by first constructing
$1-\alpha/2$ intervals for
$\mu_1$ and for~$\mu_2$, as described in the previous section, and
then calculating the minimum and maximum differences of the two
intervals. However, a more direct construction uses the
reparametrization $\nu_1=\mu_1-\mu_2$ and $\nu_2=\mu_1+\mu_2$.

Let
\[
g_{Y_1-Y_2}(\nu_1,\nu_2) = \frac{Y_1-Y_2-\nu_1}{\sqrt{h(\theta,(\nu
_2-\nu_1)/2)+h(\theta
,(\nu_2+\nu_1)/2)}}
\]
and
\[
g_{Y_1+Y_2}(\nu_1,\nu_2) = \frac{Y_1+Y_2-\nu_2}{\sqrt{h(\theta,(\nu
_2-\nu_1)/2)+h(\theta
,(\nu_2+\nu_1)/2)}},
\]
%
then $(g_{Y_1-Y_2}(\nu_1,\nu_2),g_{Y_1+Y_2}(\nu_1,\nu_2))$ has a
bivariate standard normal distribution with correlation
\begin{eqnarray*}
\rho(\nu_1,\nu_2)&=&\bigl\{h\bigl(\theta,(
\nu_2+\nu_1)/2\bigr)-h\bigl(\theta,(\nu_2-
\nu_1)/2\bigr)\bigr\} \\
&&{}/\bigl\{h\bigl(\theta,(\nu_2-
\nu_1)/2\bigr)+h\bigl(\theta,(\nu_2+\nu_1)/2
\bigr)\bigr\}.
\end{eqnarray*}
Thus,
\begin{eqnarray*}
g_{Y_1,Y_2}(\nu_1,\nu_2)&=&\bigl(g_{Y_1-Y_2}(
\nu_1,\nu_2),g_{Y_1+Y_2}(\nu_1,
\nu_2)\bigr)\\
&&{}\times\pmatrix{
1 & \rho(
\nu_1,\nu_2)
\cr
\rho(\nu_1,\nu_2) & 1 }^{-1}
\pmatrix{
g_{Y_1-Y_2}(
\nu_1,\nu_2)
\cr
g_{Y_1+Y_2}(\nu_1,\nu_2)}
\end{eqnarray*}
has a $\chi^2_{(2)}$ distribution and can serve as a pivot for
constructing confidence
regions. Specifically,
%
%
\begin{equation}
\label{CInu} C_{Y_1,Y_2}(\nu_1,\nu_2)=\bigl\{(
\nu_1,\nu_2)\dvtx g_{Y_1,Y_2}(\nu_1,
\nu_2)\le\chi_{2,1-\alpha}\bigr\}
\end{equation}
is an exact $1-\alpha$ confidence region for $(\nu_1,\nu_2)$ and, therefore,
$\{\nu_1\dvtx(\nu_1,\nu_2) \in C_{Y_1,Y_2}(\nu_1,\nu_2),
-\infty<\nu_2<\infty\}$ is a conservative confidence set for
$\nu_1$.

\subsection{An application to the iTRAQ protocol} \label{secciitraq}

In this section we analyze data from the two control experiments
mentioned in Section~\ref{secintro}. As the two biological samples in\vadjust{\goodbreak}
these experiments were identical, $\mu_{i1}-\mu_{i2}=0$ for all
peptides $i=1,\ldots,N$. In order to evaluate the performance of our
confidence intervals, we used the parameters' estimates from one
experiment to construct 95\% level confidence intervals for the
difference of
peptide abundances in the other experiment, and calculated the
proportion of intervals that did not cover 0, the true difference.

\begin{figure}

\includegraphics{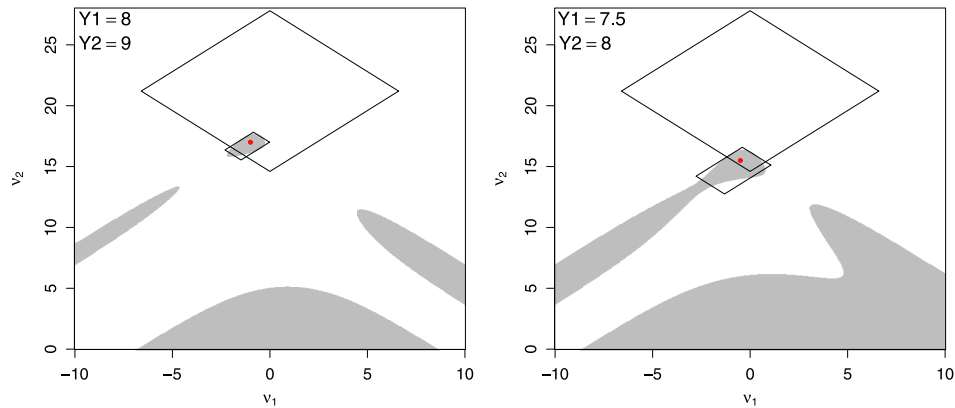}

\caption{95\% confidence regions for the pair
$(\nu_1,\nu_2)$ constructed for two data points: $(8,9)$ (left) and
$(7.5,8)$ (right).} \label{figci}
\end{figure}

As in the uni-parameter case, the confidence set for $(\nu_1,\nu_2)$
is not
necessarily a connected set and the resulting confidence set for
$\nu_1$ is not always an interval. Figure~\ref{figci}
demonstrates the shape of the confidence region for the pairs
$(Y_1,Y_2)=(8,9)$ and $(Y_1,Y_2)=(7.5,8)$. The X and Y axes are,
respectively, $\nu_1$ and $\nu_2$, and the confidence sets
are the shaded areas. The figure reveals that without
restricting the parameter space of $\mu$ (hence the parameter
space of $\nu_1$), the confidence interval for $\nu_1$ comprises all
of the
real line and is noninformative. We therefore assumed that
$\mu\in[7.3,13.9]$, where the limits were determined from
typically observed values as well as the expected range of the
intensity values. This decision restricts the values of $\nu_1$
and $\nu_2$ to the large parallelograms depicted in Figure \ref
{figci}, and
enables the construction of informative confidence intervals. The
smaller parallelograms represent confidence sets obtained by Bonferroni
correction applied to univariate confidence intervals for $\mu_1$ and
for $\mu_2$.

%
%

Using estimates from the January data, we calculated confidence
intervals for $\nu_1$ for peptides in the March experiment by (\ref
{CInu}), Bonferroni correction and the naive approach, and found that,
respectively, 18 (0.8\%), 1 (0.05\%) and 86 (3.96\%) of the 2174
intervals did not include the true parameter $\nu_1=0$. The
corresponding numbers for the 2144 peptides in the January experiments
using estimates from the March data are 47 (2.19\%), 8 (0.4\%) and 106
(4.94\%). This exercise suggests that interval (\ref{CInu}) is better
than the Bonferroni interval, but is still conservative. The naive
approach performs surprisingly well in our experiment, but, in general,
its theoretical coverage probability is not controlled. Further
research is needed to understand this phenomenon.

%
\begin{table}
\caption{95\% confidence intervals for the ratio of
phosphopeptide quantity across two experimental conditions using the
conservative and the naive approaches}
\label{tabdata}
\begin{tabular*}{\tablewidth}{@{\extracolsep{\fill}}lcd{2.2}cc@{}}
\hline
\textbf{Phosphopeptide} & \multicolumn{1}{c}{\textbf{FLT3-D835Y}}
& \multicolumn{1}{c}{\textbf{FLT3-ITD}} & \multicolumn{1}{c}{\textbf{CI}}
& \multicolumn{1}{c@{}}{\textbf{CI naive}} \\
\hline
VLPQDKEpYYK & 10.21 & 10.78 & $(0.41,0.76)$ & $(0.44,0.72)$ \\
GQESEpYGNITYPPAVR & 13.62 & 11.89 & $(5.05,6.36)$ & $(5.11,6.22)$ \\
HKEEVpYENVHSK & 11.19 & 9.92 & $(2.66,5.05)$ & $(2.76,4.59)$ \\
pYKNILPFDHSR & 10.83 & 9.80 & $(2.03,4.10)$ & $(2.12,3.69)$ \\
AVDGpYVKPQIK & 11.45 & 13.36 & $(0.13,0.17)$ & $(0.13,0.17)$ \\
\hline
\end{tabular*}
\end{table}

\subsection{An application to cancer phosphoproteomics} \label{seccicancr}

Zhang et~al. (\citeyear{Zhaetal10}) used iTRAQ labeling to study a key
modification to proteins called phosphorylation, which is
important in cell signaling, and is often deregulated in cancer
cells. Their study focused on aberrant signaling arising from
oncogenic FLT3 mutations in acute myeloid leukemia. In particular,
they monitored a key, subcomponent of signal transduction, namely,
protein tyrosine phosphorylation, in order to obtain a global
understanding of the oncogenic potential of two clinically
identified FLT3 mutants (FLT3-D835Y and FLT3-ITD). Both FLT3 mutants
induce constitutive
signaling even in the absence of proper external cues, which
results in uncontrolled cell proliferation, a~hallmark of cancer
development [\citet{BluHun01}]. Since receptor tyrosine
kinases are often constitutively active in cancer cells, it is not
surprising that a large proportion of downstream phosphorylation events
diverge from a 1:1 value (measured relative to a control cell line),
effectively eliminating the possibility of deriving a variance function
from the experimental data themselves.

Based on the mixture model results of the pooled control experiments,
we calculated 95\% confidence intervals for the ratio of
phosphopeptides mentioned explicitly in the figures of Zhang et~al.;
these are reported in Table~\ref{tabdata}. For comparison, we
calculated confidence intervals based on a naive approach mentioned in
Section~\ref{secCI2par}.
Using the intensity-dependent variance function described in this
paper, as opposed to a constant cutoff
point often used in the literature, subtle changes in
phosphorylation levels can be found (e.g., peptide VLPQDKEpYYK). Such
ratios can be considered
statistically significant despite being smaller than the typical
cutoff of 2:1. This in turn suggests that experiments can be
designed to explore phosphorylation under physiological conditions,
unlike many current experimental
designs where artificial or extreme environments are used in order
to amplify the observed ratios.

\section{Hypothesis testing}\label{secpval}

\subsection{Calculation of $p$-values}

For a given peptide, consider testing the hypothesis
$H_0\dvtx\mu_1=\mu_2=\mu$ versus the two-sided alternative
$H_1\dvtx\mu_1\neq\mu_2$. Theoretically, this can be done by inverting
the confidence intervals discussed in the previous section. However,
for calculating $p$-values, this inversion is computationally difficult
and the current section explores an alternative approach.

As before, let $Y_1\sim N(\mu_1,h(\theta,\mu_1))$ and $Y_2\sim
N(\mu_2,h(\theta,\mu_2))$ be independent, then $Y_1-Y_2\sim N(\mu_1-\mu
_2,h(\theta,\mu_1)+h(\theta,\mu_2))$ and a reasonable test may compare
%
%
\begin{equation}
\label{eqtest} \frac{(Y_1-Y_2)^2}{2h(\theta,\mu)}
\end{equation}
to the chi-squared distribution with one degree of freedom. However,
(\ref{eqtest}) contains the unknown parameters $\theta$ and $\mu$, and
hence is not a legitimate test statistic. As in the construction of
confidence intervals, a naive $p$-value can be calculated by replacing
the denominator of (\ref{eqtest}) with $2h(\hat{\theta},(Y_1+Y_2)/2)$.
Although it is reasonable to replace $\theta$ with its consistent
estimator $\hat{\theta}$, the average $(Y_1+Y_2)/2$ is an inconsistent
estimator for $\mu$, possibly leading to an anti-conservative $p$-value.

As in other statistical problems involving nuisance parameters, an
asymptotically conservative $p$-value is obtained by
\[
\sup_{\mu\in[a,b]} P\bigl(Z^2>(y_1-y_2)^2/2h(
\hat{\theta},\mu)\bigr), 
\]
where $y_1$ and $y_2$ are the realized values and $Z^2$ is a $\chi
^2_{(1)}$ random variable.\vspace*{1pt}
%
%
This approach is simple and easy to implement but results in an overly
conservative $p$-value. We therefore suggest to employ the approach of
\citet{BerBoo94}, where the $p$-value is calculated by maximization
over a confidence interval for the nuisance parameter.
Specifically, let $C_\beta$ be a $1-\beta$ level confidence interval
for~$\mu$, obtained in a way similar to that presented in Section \ref
{secCI1par}, then we define our $p$-value as
%
\begin{equation}
\label{eqbb} \sup_{\mu\in C_\beta} P\bigl(Z^2>(y_1-y_2)^2/2h(
\hat{\theta},\mu)\bigr)+\beta.
\end{equation}
The choice of $\beta$ depends on the context. If a univariate
hypothesis is tested with a significant level of 5\%, then $\beta
=0.001$ is usually a good choice. However, if a Bonferroni correction
is needed, then $\beta$ must be much smaller, as $p\mbox{-value}>\beta$ by
construction.

%
%
%
%
\begin{figure}

\includegraphics{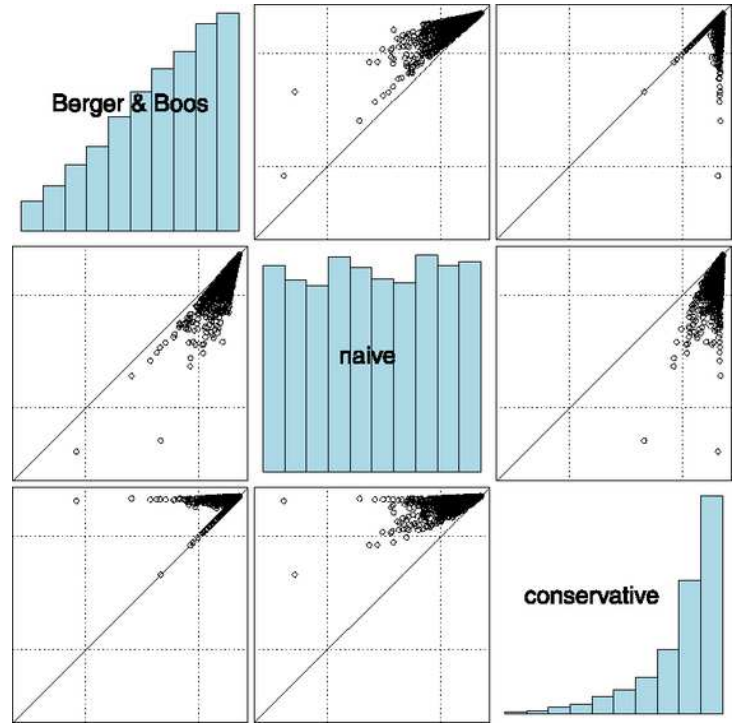}

\caption{Scatterplots and histograms of $p$-values for the pooled control
experiments obtained by different approaches. Scatterplots are in log
scale with dotted lines indicating significant level of 5\% with and
without Bonferroni correction.} \label{figpval}
\end{figure}
%

\subsection{Application to the iTRAQ data}

$p$-values for testing no difference of peptide amounts were calculated
for all pairs in the pooled iTRAQ control experiments described in
Section~\ref{secciitraq}. We used the variance function $h(\theta,\mu
)=\exp(\theta_1+\theta_2\mu)$ with $\theta$ estimated by the EM\vadjust{\goodbreak}
algorithm applied to the pooled data. Three approaches were compared:
the naive approach that replaces $\mu$ in $h(\theta,\mu)$
with $\bar {Y}=(Y_1+Y_2)/2$, a conservative simple approach that
replaces\vspace*{1pt} $\mu$ with $a$, and the approach of Berger and Boos based on
(\ref{eqbb}) with $\beta=10^{-6}$.

Figure~\ref{figpval} presents scatterplots and histograms of the
$p$-values. Scatterplots are depicted in logarithmic scale with dotted
lines indicating 0.05 and $0.05/N$ significant levels. Recall that the
null hypothesis holds in these experiments. Two peptides had naive
$p$-values smaller than the Bonferroni cutoff value, for one of them the
Berger and Boos $p$-value was also under the cutoff level. For about 40\%
of the experiments the Berger and Boos $p$-value was similar to the
conservative $p$-value, but for the other peptides, the Berger and Boos
method gives considerable smaller values. These are valid $p$-values and
certainly worth the additional computation effort. The distribution of
$p$-values under the naive approach is very close to the uniform
distribution, whereas the distributions under the other two approaches
are stochastically larger. This and the simulation presented in the
next section suggest that the naive approach for testing differences of
peptide amounts may be only slightly anti-conservative, though a more
extensive study is required before the naive approach can be recommended.

As a second application, we calculated $p$-values for the data described
in Section~\ref{seccicancr} that contrast two mutants found often in
certain types of cancer. There are $N=205$ peptides in the data, the
$p$-values of 141, 123 and 9 of them were smaller then $0.05/N$ according
to the naive, the Berger and Boos and the conservative approach,
respectively. These peptides are then prioritized for in-depth
functional characterization.

\section{Simulation} \label{secsim}

\subsection{Estimating the variance function}

The performances of the two estimation approaches, the MACL and the
mixture model, were
tested by simulation under various conditions. The first set of
simulations aimed at testing the performance of the MACL
approach. For each of the scenarios described below, we generated
values for the nuisance parameters and then generated independent
pairs of observations from the corresponding normal distributions.
We repeated this process 1000 times for different sample sizes
($N=200, 500, 1000$ and 2000) and for two different variance
functions of the form $\exp(\theta_1+\theta_2\mu)$:
$(\theta_1,\theta_2)=(5,-1)$, which is similar to the values
obtained in our data, and $(\theta_1,\theta_2)=(5,-0.5)$, which
reflects observations with a much larger variance. The following
scenarios were considered:
\begin{itemize}
\item Observations fixed: a set of $\mu_i$ values was sampled from the
observed $\bar{Y}_i$'s (with replacement) and the same values
were used in all 1000 replications.
\item Observations random: a different set of $\mu_i$ values was sampled
from the observed $\bar{Y}_i$'s (with replacement) for each
simulation.
\item U$(8,12)$: the $\mu_i$ values were generated from the
continuous uniform distribution over $(8,12)$.
\item U$\{8,9,\ldots,12\}$: the $\mu_i$ values were generated from the
discrete uniform distribution over $\{8,9,\ldots,12\}$.
\end{itemize}

The results of the simulation are summarized in Table
\ref{tabsimMACL} and in more details in the supplemental article
[\citet{Manetal}]. There seems to be
almost no difference between the scenarios considered (see Table 1 of
the supplementary materials), and this suggests that the
approach is insensitive to modest changes in the distribution of
the nuisance parameters. Both the variance and the bias decrease with
sample size for the model $(\theta_1,\theta_2)=(5,-1)$, and the
overall performance of the MACL approach for this case is
satisfactory. However, the MACL estimators are biased for the
case $(\theta_1,\theta_2)=(5,-0.5)$, and the bias did not decrease
with sample size (Table~\ref{tabsimMACL}). Thus, unless the variance is
very small, the approach is problematic and is not recommended.

%
\begin{table}
\def\arraystretch{0.9}
\tablewidth=350pt
\caption{Simulation results for the MACL method under the observations
fixed scenario}\label{tabsimMACL}
\begin{tabular*}{\tablewidth}{@{\extracolsep{\fill}}rcd{2.3}cd{2.1}cc@{}}
\hline
& \multicolumn{3}{c}{$\bolds{\theta_1}$} &
\multicolumn{3}{c}{$\bolds{\theta_2}$}\\[-4pt]
& \multicolumn{3}{c}{\hrulefill} & \multicolumn{3}{c@{}}{\hspace*{-1.5pt}\hrulefill}\\
\multicolumn{1}{@{}l}{$\bolds{N}$} & \multicolumn{1}{c}{$\bolds{\theta_1}$}
& \multicolumn{1}{c}{\textbf{bias}} & \multicolumn{1}{c}{\textbf{std}}
& \multicolumn{1}{c}{$\bolds{\theta_2}$} & \textbf{\textbf{bias}}
& \multicolumn{1}{c@{}}{\textbf{std}} \\
\hline
200 & 5 & -0.061 & 0.830 & -1 & 0.006 & 0.081\\
500 & 5 & -0.051 & 0.508 & -1 & 0.005 & 0.049 \\
1000 & 5 & -0.020 & 0.360 & -1 & 0.002 & 0.036 \\
2000 & 5 & -0.002 & 0.251 & -1 & 0.000 & 0.025 \\
[6pt]
200 & 5 & -1.238 & 0.759 & -0.5 & 0.127 & 0.074 \\
500 & 5 & -1.175 & 0.484 & -0.5 & 0.121 & 0.047 \\
1000 & 5 & -1.156 & 0.319 & -0.5 & 0.119 & 0.031 \\
2000 & 5 & -1.164 & 0.238 & -0.5 & 0.120 & 0.023 \\
\hline
\end{tabular*}
\end{table}

In order to test the mixture model approach, we generated $\mu_i$
by the observations fixed scenario under the two variance functions
described above. For each sample size, we simulated 200 data sets
and calculated the empirical biases and standard deviations. The
results are listed in Table~\ref{tabsimMIX}. As expected, the
bias and variance of the estimators decrease with sample size for
both models.

%
\begin{table}[b]
\def\arraystretch{0.9}
\tablewidth=350pt
\caption{Simulation results for the mixture model method under the
observations fixed scenario}\label{tabsimMIX}
\begin{tabular*}{\tablewidth}{@{\extracolsep{\fill}}rcd{2.3}cd{2.1}d{2.3}c@{}}
\hline
& \multicolumn{3}{c}{$\bolds{\theta_1}$} & \multicolumn{3}{c@{}}{$\bolds{\theta_2}$}\\[-4pt]
& \multicolumn{3}{c}{\hrulefill} & \multicolumn{3}{c@{}}{\hspace*{-1.5pt}\hrulefill}\\
\multicolumn{1}{@{}l}{$\bolds{N}$} &
\multicolumn{1}{c}{$\bolds{\theta_1}$} & \multicolumn{1}{c}{\textbf{bias}}
& \multicolumn{1}{c}{\textbf{std}}
& \multicolumn{1}{c}{$\bolds{\theta_2}$} & \multicolumn{1}{c}{\textbf{bias}}
& \multicolumn{1}{c@{}}{\textbf{std}} \\
\hline
200 & 5 & 0.580 & 0.914 & -1 & -0.071 & 0.088 \\
500 & 5 & 0.423 & 0.534 & -1 & -0.049 & 0.052 \\
1000 & 5 & 0.274 & 0.328 & -1 & -0.030 & 0.032 \\
2000 & 5 & 0.173 & 0.227 & -1 & -0.019 & 0.022 \\
[6pt]
200 & 5 & 0.070 & 1.026 & -0.5 & -0.009 & 0.099 \\
500 & 5 & 0.019 & 0.608 & -0.5 & -0.003 & 0.059 \\
1000 & 5 & 0.027 & 0.397 & -0.5 & -0.003 & 0.039 \\
2000 & 5 & -0.013 & 0.291 & -0.5 & 0.001 & 0.028 \\
\hline
\end{tabular*}
\end{table}

\begin{figure}

\includegraphics{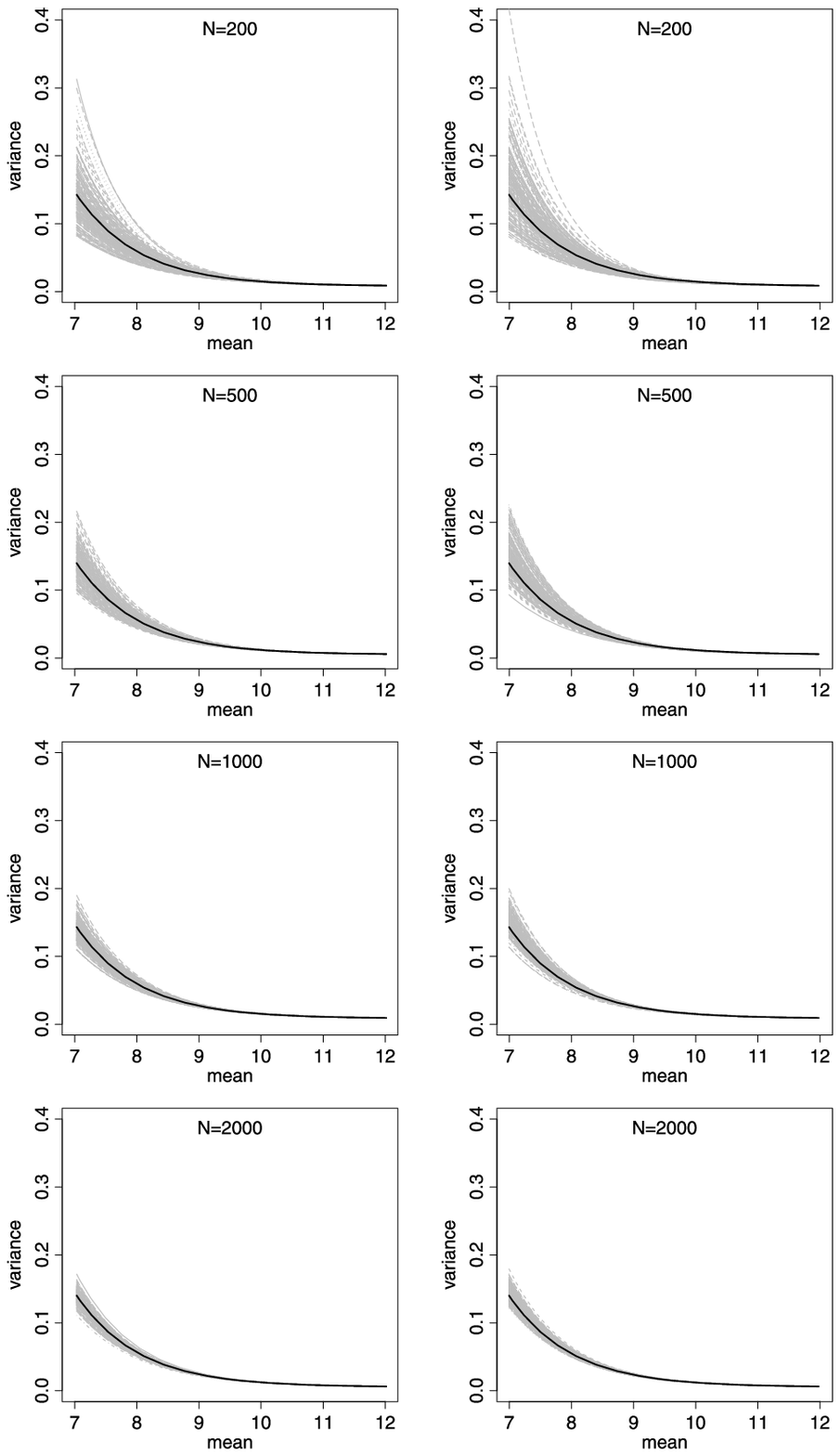}

\caption{Estimated variance functions (gray) and the
true variance function (black) obtained in 200 simulated data sets
for the case $(\theta_1,\theta_2)=(5,-1)$. The figures on the left
show the results of the MACL approach and those on the right are for
the mixture
model approach. The simulated data sample sizes are, from top to
bottom, 200, 500, 1000 and 2000.} \label{figsim1}
\end{figure}

\begin{figure}

\includegraphics{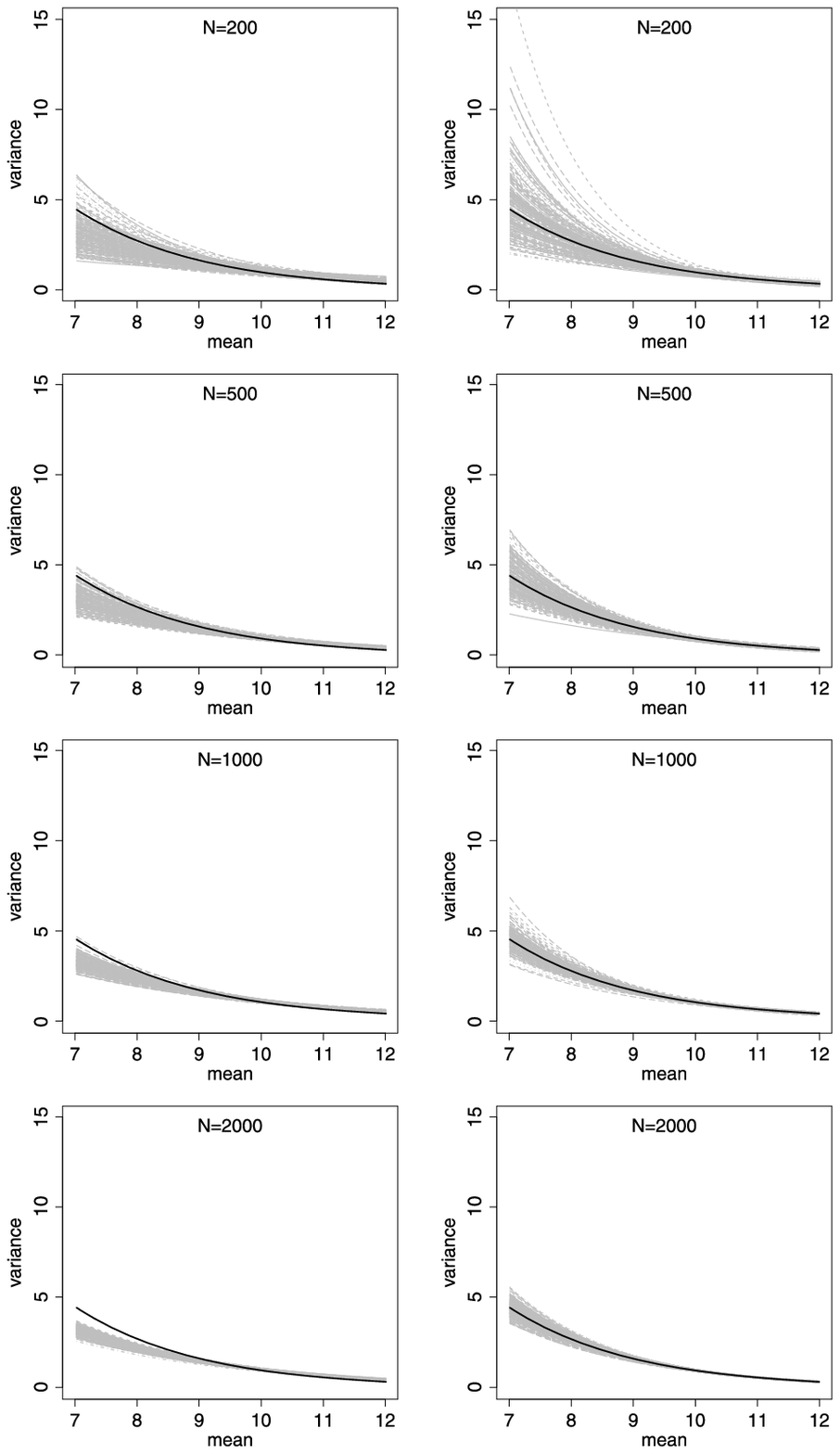}

\caption{Estimated variance functions (gray) and the
true variance function (black) obtained in 200 simulated data sets
for the case $(\theta_1,\theta_2)=(5,-0.5)$. The figures on the
left show the results of the MACL approach and those on the right are
for the mixture
model approach. The simulated data sample sizes are, from top to
bottom, 200, 500, 1000 and 2000.} \label{figsim2}
\end{figure}

Figures~\ref{figsim1} and~\ref{figsim2} display the
performance of the estimators for the variance, that is, the
performance of $\exp(\hat{\theta}_1+\hat{\theta}_2\mu)$ as a
function of $\mu$. The gray lines are estimated variance functions
%
%
%
%
from 200 simulated data sets and the true variance
function is depicted in black. The figures demonstrate that the mixture
model approach is as good as the MACL in the low variance case and
performs better in the large variance scenario, especially when the
sample size is large.

\subsection{\texorpdfstring{Confidence intervals for $\mu$}{Confidence intervals for mu}}

Intervals for $\mu$ based on one observation use the pivot $(Y-\mu
)^2/h(\theta,\mu)$ and are exact. Here we study the performance of the
corresponding $1-\alpha$ naive confidence intervals that replace $\mu$
with $Y$ in the variance function $h(\theta,\mu)$; see Section \ref
{secCI1par}.

%
\begin{figure}

\includegraphics{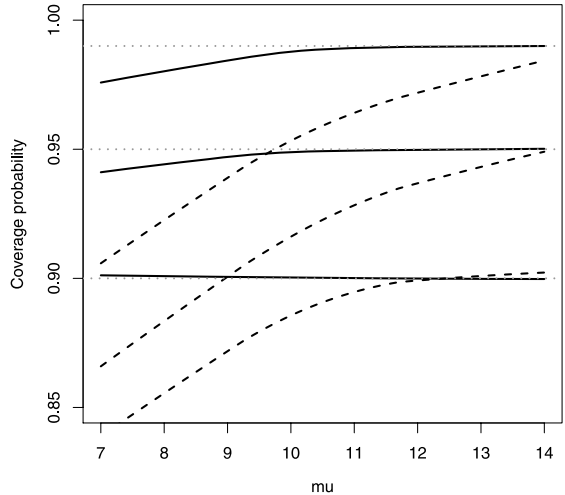}

\caption{Coverage probability of the naive confidence interval for
$\mu
$ for confidence levels 0.99, 0.95 and 0.90. Solid lines: $h(\theta
,\mu
)=\exp(5-\mu)$, dashed lines: $h(\theta,\mu)=\exp(5-0.5\mu)$.}
\label{figsimCI}
\end{figure}

Figure~\ref{figsimCI} presents the coverage probability of the naive
intervals as a function of the mean ($\mu=7, 7.1,\ldots,14$), the coverage
probability ($1-\alpha=0.9, 0.95, 0.99$) and the variance function
[$\exp(5-\mu)$, $\exp(5-0.5\mu)$]. For each $\mu$, $\alpha$ and
variance function $h(\theta,\mu)$, we estimated the coverage
probability by simulating 100,000 replications from the model $N(\mu
,h(\theta,\mu))$, constructing naive confidence intervals of level
$1-\alpha$, and calculating the proportion of intervals covering $\mu$.
The performance depends on the variance at $\mu$, where the true
coverage for small $\mu$ (i.e., a large variance) could be much lower
than the aimed coverage, especially for large values of $1-\alpha$ and
for the model $h(\theta,\mu)=\exp(5-0.5\mu)$ represented by dashed lines.

\subsection{$p$-values}

A third simulation study was conducted with the aim of understanding
the properties of $p$-values obtained by the different approaches. For a
selected set of values for $\mu$, we generated 10,000 independent pairs
of observations $(Y_1,Y_2)$ such that $Y_1\sim N(\mu,e^{\theta
_1+\theta
_2\mu})$ and $Y_2\sim N(\mu_k,e^{\theta_1+\theta_2\mu_k})$,
independently, where $\mu_k=\mu+k\times\sqrt{e^{\theta_1+\theta
_2\mu
}}$, that is, $Y_1$ and $Y_2$ are centered about $k$ standard
deviations apart. The parameter $k$ ranged from 0 to 3, and for
$(\theta_1,\theta_2)$ we studied the values $(5,-1)$ and $(5,-0.5)$. We used
the value $\beta=10^{-3}$ for the Berger and Boos $p$-values.

Figures 4 and 5 of the supplemental article [\citet{Manetal}]
present the proportions of $p$-values that were smaller than 0.05 as a
function of $\mu$, $k$ and~$\theta_2$. For the case $\theta_2=-1$
(supplemental article, Figure 5), the naive approach is only slightly
anti-conservative\vadjust{\goodbreak} and only for small values of $\mu$, and its power is
much larger than that of the other methods. The Berger and Boos
approach works reasonably well only for very small and very large
values of $\mu$; the conservative approach is useless. For the case
$\theta_2=-0.5$ (supplemental article, Figure 4), the naive approach
is anti-conservative for small values of $\mu$ with a significant level
of up to 0.08 instead of the declared level of 0.05. However, it is the
only method that has a useful power function.\looseness=1

\section{Discussion} \label{secdisc}

This paper presents a protocol for estimating the variance
function of a mass spectrometer when used for relative quantification
in proteomic applications. Using two sets of data collected
three months apart, we found that the variance function is stable
over time. However, we expect to find different variance functions
in different instruments and, hence, each lab should estimate
the variance parameters of each spectrometer independently using the protocol
described here, and update them periodically. More importantly, the
variance estimated here corresponds only to the variance of the
instrument itself, and does not include error terms corresponding to
the natural variability of biological samples or the processing
required to solubilize proteins from cells and tissues, generate
peptides, etc.

Two inference approaches are considered. The first estimates the
nuisance parameters by simple averages and then maximizes a target
function; the second approach assumes a mixture model and estimates the
variance function by maximum likelihood. When the variance is small and
changes slowly as a function of the mean, as is the case in the data we
analyzed, an average of two iTRAQ reporter ions per peptide provides a
reasonable estimate for the unknown $\mu$ and the first approach gives
good results. However, it may yield highly biased estimators in other
scenarios, as demonstrated by simulations, and we therefore recommend
the routine use of the mixture model approach that has a sound
theoretical justification.

When using the estimated variance function for statistical inference on
one parameter $\mu$, exact methods for constructing confidence
intervals are available and are much more appropriate than intervals
constructed by the anti-conservative naive approach. On the other hand,
for inference on the difference or the ratio of peptide abundance
measured across two biological conditions, the naive approach performs
quite well, yielding a significant level only slightly larger than the
aimed one.

The iTRAQ protocol is somewhat more complicated than presented
here, as it involves preprocessing of the spectral data to correct
for differences in total protein amount, iTRAQ label purity and
instrument-specific parameters. Moreover, iTRAQ is known to suffer from
contamination due to co-eluting chromatographic peaks that also share
similar precursor masses (i.e., peptides which have a
similar $m/z$ value and retention time). Theoretically, these factors may induce
dependence between measurements that is ignored in the current
analysis.

In order to test the underlying normal assumption, Zhang et~al.
produced a q--q plot of
$(Y_{i1}-Y_{i2})/\sqrt{h(\hat{\theta},\bar{Y}_i)}$ that showed a
very good fit. Although the graph is suggestive, it relies on
$\bar{Y}_i$ as an estimate for the nuisance parameter $\mu_i$,
hence it does not have a theoretical support. A~formal approach we
intend to explore requires at least three observations for each
peptide.
A~simple transformation of each of the triplets results in
variables that, under the normal model, have a Cauchy
distribution, and a q--q plot or formal goodness-of-fit tests can
be easily employed. This approach requires iTRAQ data from a control
experiment that yields more than two independent
and identically distributed measures for each peptide. We intend to
conduct such an experiment in the future.

The conservative results of the exercise conducted in Section
\ref{secCI2par} are partially due to the inherently conservative
construction of the intervals, but may also be a result of the
special parametric shape for the variance function we considered.
Our chosen parametric model is very simple, enabling the
implementation of simple algorithms, and is supported quite well
by the data. A nonparametric method has been recently suggested
for the analysis of microarray data [\citet{CarWan08}]; the
possibility of adopting this method to MS data and of using it for
goodness-of-fit testing should be further explored.

\begin{appendix}\label{app}
\section{Bias of the estimating equations}\label{appA}

Recall that $\bar{Y}_i\sim
N(\mu_i,\frac{1}{2}e^{\theta_1+\theta_2\mu_i})$ and $S_i^2\sim
e^{\theta_1+\theta_2\mu_i}\chi^2_{(1)}$, and that $\bar{Y}_i$ and
$S_i^2$ are independent. Straightforward calculations show that
$E(S_i^2)=e^{\theta_1+\theta_2\mu_i}$ and
$E\{\exp(-\theta_1-\theta_2
\bar{Y}_i)\}=\exp(-\theta_1-\theta_2\mu_i+\frac{1}{4}\theta_2^2
e^{\theta_1+\theta_2\mu_i})$ and, therefore,
\[
E\Biggl\{N^{-1}\sum_{i=1}^N
S_i^2\exp(-\theta_1-\theta_2
\bar{Y}_i)\Biggr\}= N^{-1}\sum_{i=1}^N
\exp\biggl(\frac{1}{4}\theta_2^2e^{\theta_1+\theta
_2\mu_i}
\biggr)
\]
so that the expectation of the first equation (\ref{eqee1}) differs
from zero,
unless $\theta_2=0$, that is, the homogeneous model of \citet{NeySco48}.


For the second equation, we have $E\bar{Y}_i
\exp({-\theta_2\bar{Y}_i})=-\frac{d}{dt}E
\exp({-t\bar{Y}_i})|_{t=\theta_2}= -\frac{d}{dt}
\exp\{-t \mu_i+
t^2\frac{1}{4}e^{\theta_1+\theta_2\mu_i}\}|_{t=\theta_2}= (\mu_i-
\frac{1}{2}\theta_2e^{\theta_1+\theta_2\mu_i})\exp\{-\theta_2
\mu_i+ \theta_2^2\frac{1}{4}\times\break e^{\theta_1+\theta_2\mu_i}\}$, so that
$
E\{ \bar{Y}_i S_i^2\exp(-\theta_1-\theta_2
\bar{Y}_i)\}= (\mu_i-
\frac{1}{2}\theta_2e^{\theta_1+\theta_2\mu_i})
\exp(\frac{1}{4}\theta_2^2\times e^{\theta_1+\theta_2\mu_i}),
$
and the expectation of the left-hand side of (\ref{eqee2}) is
\[
N^{-1}\sum_{i=1}^N
\mu_i - N^{-1}\sum_{i=1}^N
\biggl(\mu_i- \frac{1}{2}\theta_2e^{\theta_1+\theta_2\mu_i}
\biggr) \exp\biggl(\frac{1}{4}\theta_2^2e^{\theta_1+\theta_2\mu_i}
\biggr),
\]
which again differs from 0 for $\theta_2\neq0$.

In general, the bias will be small if $e^{\theta_1+\theta_2\mu_i}$
is small for all $i$, which means that the variance of the
measurements is small and, hence, the local averages are good
estimators for the unknown $\mu_i$ parameters.\vspace*{-2pt}

\section{Consistency of the MLE}\label{appB}

A generic sample point is $y=(y_1,y_2)$, where, by conditional independence,
%
%
\begin{equation}
\label{eqbinorm} f(y;\theta|\mu)=\bigl\{2\pi h(\theta,\mu)\bigr\}^{-1}
\exp\biggl\{-\frac{(y_1-\mu)^2+(y_1-\mu)^2}{2h(\theta,\mu)} \biggr\}.
\end{equation}
The marginal density of $Y=(Y_1,Y_2)$ is $g(y;\theta,G_0)=\int_t
f(y;\theta|t)\,dG_0(t)$.

The proof of consistency is based on the result of
\citet{KieWol56} (KW hereafter); the metric we use below is given
in KW equation~(2.2).

We complete the parameter space of $(\theta,G_0)$ by including all
proper distributions functions with support in $[a,b]$.

Assumption 1 of KW trivially holds with respect to the Lebesgue
measure. Next, note that $f(y;\theta|\mu)$, and hence $g(y;\theta
,G_0)$, is bounded above by $\alpha^{-1}$. Let $(\theta
_i,G_i)\rightarrow(\theta^*,G^*)$, where $(\theta^*,G^*)$ is in the
complete parameter space. In order to verify
Assumption 2 of KW, we need to show that $g(y;\theta_i,G_i)
\rightarrow
g(y;\theta^*,G^*)$. We have
\begin{eqnarray*}
&& \bigl|g(y;\theta_i,G_i) - g\bigl(y;\theta^*,G^*\bigr) \bigr|
\\[-2pt]
&&\qquad= \biggl|\int f(y;\theta_i|t)\,dG_i(t) - \int f\bigl(y;
\theta^*|t\bigr)\,dG^*(t) \biggr|
\\[-2pt]
&&\qquad \le\biggl|\int\bigl\{f(y;\theta_i|t) - f\bigl(y;\theta^*|t\bigr)\bigr
\}\,dG^*(t) \biggr| + \biggl|\int f(y;\theta_i|t)\,d\bigl(G_i(t)-G^*(t)
\bigr) \biggr|
\\[-2pt]
&&\qquad \le\int\bigl|f(y;\theta_i|t) - f\bigl(y;\theta^*|t\bigr) \bigr|\,dG^*(t) +
\alpha^{-1}\int d\bigl|G_i(t)-G^*(t)\bigr|.
\end{eqnarray*}
The first term vanishes by the Dominated Convergence theorem and the
second vanishes by the convergence of $G_i$ to $G^*$.

For verifying Assumption 3 of KW, define $m(y;\theta^*,G^*,\rho)=\sup
g(y;\theta,G)$, where the supremum is taken over all $(\theta,G)$ such
that $|\theta-\theta^*|+|G-G^*|<\rho$. We need to show that $m$ is a
measurable function of $y$ for any $\rho>0$ and any $(\theta^*,G^*)$ in
the complete parameter space. This is true for the same arguments given
by KW in their first example: $g$ is for each $y$ continuous in
$(\theta
,G)$ and the parameter space is separable. To show this formally,
define $A(\theta^*,G^*,\rho,c)=\{y\dvtx m(y;\theta^*,G^*,\rho)>c\}$,
and let
$\{(\theta_i,G_i)\}$ and $\{y_j\}$ be dense subsets in the parameter
and sample space, respectively. Let $B(y,r)$ and $B(\theta,G,r)$ be
balls of radius $r$ around the corresponding points, then\vspace*{-1pt}
\[
A\bigl(\theta^*,G^*,\rho,c\bigr)=\bigcap_{n=1}^\infty
\bigcup_{j}B(y_j,1/n),\vspace*{-1pt}
\]
where the union is over \mbox{$\{j\dvtx\exists(\theta_i,G_i)\in B(\theta
^*,G^*,\rho)
\mbox{ such that }g(y_j;\theta_i,G_i)>c\}$}.

Assumption 4 of identification follows from Bruni and Koch
[(\citeyear{BruKoc85}), Theorem~1],
that proves that $G_0$ and $h(\theta,\mu)$ are identifiable
on the support of~$\mu$. Assumption (iii) below equation (\ref{eqmixmod})
ensures identifiability of $\theta$.\vadjust{\goodbreak}

To verify Assumption 5, note that our assumptions on $h(\theta,\mu)$
guarantee that $g(y;\theta,G)$ is bounded above and below so that
$E\log
\{g(Y;\theta,G)\} > -\infty$, where the expectation is taken with
respect to $g(y;\theta_0,G_0)$, the true density of $Y$.\vspace*{-3pt}
%

\section{An EM algorithm} \label{appC}\label{secem}\vspace*{-1pt}

Let $f(y;\theta|\mu)$ be the bivariate normal density defined in
(\ref{eqbinorm}), and let $a\le\mu_1\le\cdots,\mu_J\le b$ be
fixed scalars
(support points of $G_0$). We approximate the likelihood of one
observed pair by the following discrete mixture model:\vspace*{-1pt}
%
%
\begin{equation}
g(y;\theta,{\bolds\pi})=\sum_{j=1}^J
\pi_j f(y;\theta|\mu_j),\vspace*{-1pt}
\end{equation}
where ${\bolds
\pi}=(\pi_1,\ldots,\pi_J)$, $\pi_j \ge0$ and $\pi_1+\cdots+\pi_J=1$.
The unknown
parameters are the $\pi_j$'s and $\theta$.

To construct the EM algorithm, consider a Multinomial variable
$\Delta$ over $1,\ldots,J$ with a probability vector ${\bolds
\pi}$, and define
$(\delta_1,\ldots,\delta_J)$ by $\delta_j=I\{\Delta=j\}$, where
$I$ is the indicator function. Let
$Y=(y_1,\ldots,y_N)$ be data on $N$ pairs, then the complete
log likelihood can be written as\vspace*{-1pt}
%
%
\begin{equation}
\label{eqcomplik} \ell({\bolds\pi},\theta;Y) = \sum
_{i=1}^N\sum_{j=1}^J
\delta_{ij} \log\bigl\{f(y_i;\theta|\mu_j)
\bigr\} + \sum_{i=1}^N\sum
_{j=1}^J \delta_{ij}\log(
\pi_j),\vspace*{-1pt}
\end{equation}
where $\delta_{ij}$ is an indicator for the (unobserved) event \{pair
$i$ has mean $\mu_j$\}.

Denote by $\mathit{old}$ the current estimates of the unknown parameters,
then, using the Bayes formula, the expectation step reduces to
estimating\vspace*{-1pt}
\begin{eqnarray*}
E^{\mathrm{old}}(\delta_{ij}|Y)&=&E^{\mathrm{old}}(\delta_{ij}|y_i)
= P^{\mathrm{old}}(\delta_{ij}=1|y_i) \\[-2pt]
&=& \frac{\pi_j^{\mathrm{old}}
f(y_i;\theta^{\mathrm{old}}|\mu_j)} {\sum_{k=1}^J \pi_k^{\mathrm{old}}
f(y_i;\theta^{\mathrm{old}}|\mu_k)}
=: w^{\mathrm{old}}_{ij}.\vspace*{-1pt}
\end{eqnarray*}
Note that $\sum_{j=1}^J \delta_{ij}=1$ by definition, so the above
formula can be interpreted as the current estimate of the probability
that $y_i$ was
generated by the distribution having mean $\mu_j$.

The maximization step is obtained by replacing $\delta_{ij}$ in (\ref
{eqcomplik}) with $w^{\mathrm{old}}_{ij}$ and solving\vspace*{-1pt}
%
%
\begin{equation}
\max_{{\bolds\pi},\theta} \sum_{i=1}^N\sum
_{j=1}^J w^{\mathrm{old}}_{ij}
\log\bigl\{f(y_i;\theta|\mu_j)\bigr\} + \sum
_{i=1}^N\sum_{j=1}^J
w^{\mathrm{old}}_{ij}\log(\pi_j),\vspace*{-1pt}
\end{equation}
which is done separately for ${\bolds\pi}$ and $\theta$.
For $\theta$, the problem is of a nonparametric regression type
and can be solved by reweighted least squares, similar to the MACL
approach. The mixing
probabilities are simply updated by\vspace*{-1pt}
\[
\pi_j^{\mathrm{new}} = \frac{1}{N} \sum
_{i=1}^N w^{\mathrm{old}}_{ij}.\vspace*{-1pt}
\]
\end{appendix}\eject


\section*{Acknowledgments}

We thank Yosi Rinott for helpful comments regarding the proof of
Theorem~\ref{theo1}. We thank the referees and the Associate Editor for many
helpful comments and suggestions.

\begin{supplement}
\stitle{Web-based supplementary materials
variance function estimation in quantitative mass
spectrometry with application to iTRAQ labeling\\}
\slink[doi,text={10.1214/ 12-AOAS572SUPP}]{10.1214/12-AOAS572SUPP} 
\sdatatype{.pdf}
\sfilename{aoas572\_supp.pdf}
\sdescription{Section A: Workflow of the iTRAQ technique. Section~B:
Estimate of
$G_0$. Section C: Sensitivity of the EM algorithm to initial values.
Section D: Detailed simulation results.}
\end{supplement}

%

\printaddresses

\end{document}